\documentclass{phbauth}

\usepackage{graphicx,psfig,amssymb}
\usepackage[figuresright]{rotating}

\begin{document}

\begin{frontmatter}

  \title{Exponents of the localization lengths \\ in the bipartite
    Anderson model with off-diagonal disorder}

\author[JUK]{Andrzej Eilmes\thanksref{AE}},
\author[TUC]{Rudolf A.\ R\"{o}mer\thanksref{RAR}} and 
\author[TUC]{Michael Schreiber\thanksref{CA}}

\address[JUK]{Department of Computational Methods in Chemistry,
  Jagiellonian University, Ingardena 3, 30-060 Krak\'{o}w, Poland}

  \address[TUC]{Institut f\"{u}r Theoretische Physik, Technische
    Universit\"{a}t, 09107 Chemnitz, Germany}

\thanks[AE]{E-mail: eilmes@chemia.uj.edu.pl}
\thanks[RAR]{E-mail: r.roemer@physik.tu-chemnitz.de}
\thanks[CA]{Corresponding author.\\\mbox{   } E-mail: schreiber@physik.tu-chemnitz.de}

\begin{abstract}
  We investigate the scaling properties of the two-dimensional (2D)
  Anderson model of localization with purely off-diagonal disorder
  (random hopping). In particular, we show that for small energies the
  infinite-size localization lengths as computed from
  transfer-matrix methods together with finite-size scaling diverge
  with a power-law behavior. The corresponding exponents seem to
  depend on the strength and the type of disorder chosen.\vspace{1pc}
\end{abstract}

\end{frontmatter}

\maketitle


\section{Introduction}
\label{sec-intro}

Of paramount importance for the theory of disordered systems and the
concept of Anderson localization
\cite{And58,EcoC70,EcoC72,Eco72,LicE74} is the scaling theory of
localization as proposed in 1979 \cite{AbrALR79}. Especially in 2D,
this theory predicts the absence of a disorder-driven MIT for generic
situations such that all states remain localized and the system is an
insulator \cite{LeeR85,KraM93,BelK94}.  However, already early
\cite{EcoA77,AntE77} it was suggested that an Anderson model of
localization with purely off-diagonal disorder might violate this
general statement since non-localized states were found at the band
center \cite{Oda80,SouE81,FerAE81,PurO81,SouWGE82}.  Further numerical
investigations in recent years
\cite{EilRS98a,EilRS98b,Cai98T,CaiRS99,BisCRS99} have uncovered
additional evidence that the localization properties at $E=0$ are
special. In particular, it was found that the divergence in the
density of states DOS is accompanied by a divergence of the
localization lengths $\lambda$ \cite{EilRS98a,EilRS98b}. This
divergence does not violate the scaling arguments \cite{GadW91}, since
it can be shown that its scaling properties are compatible with
critical states only \cite{EilRS98b}, i.e., there are no truly
extended states at $E=0$. Of importance for the model is a very
special symmetry around $E=0$ which holds in the bipartite case of an
even number of sites \cite{GadW91,Gad93}.  Then the spectrum is
symmetric such that for every eigenenergy $E_{i}<0$ there is also a
state with energy $E_{i}>0$. This situation is connected with a
so-called chiral universality class.  Furthermore, the model is
closely connected to the random flux model studied in the quantum-Hall
situation where the off-diagonal disorder is due to a random magnetic
flux through the 2D plaquettes.

Thus although we do not have a true MIT, we nevertheless have a
transition from localized via delocalized to localized behavior as we
sweep the energy through $E=0$. We consider a single electron on the
2D lattice with $N$ sites described by the Anderson Hamiltonian
\begin{equation}
  H = \sum_{i \neq j}^N t_{ij} \left| i \right\rangle \left\langle j
  \right| + \sum_i^N \epsilon_i \left| i \right\rangle \left\langle i
  \right|
\label{hamilt}
\end{equation}
where $\left| i \right\rangle$ denotes the electron at site $i$. The
onsite energies $\epsilon_i$ are set to $0$ and the off-diagonal
disorder is introduced by choosing random hopping elements $t_{ij}$
between nearest neighbor sites. 

\begin{figure}[tb]
\vspace{0pt}
\centerline{\psfig{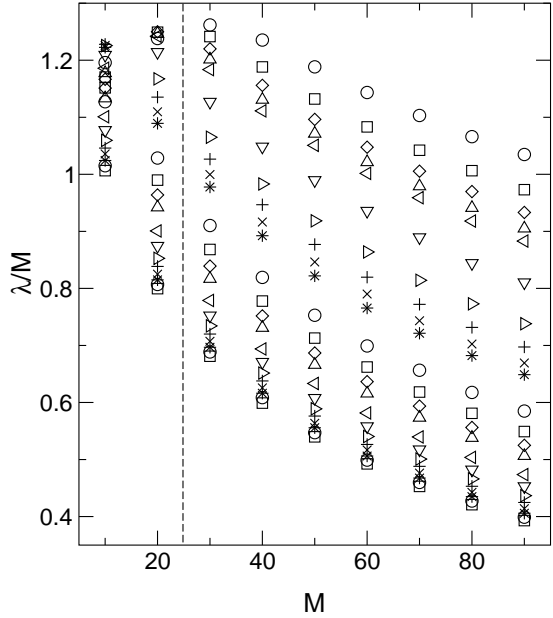}}
\caption{\label{fig-c000B}
  Reduced localization length $\lambda/M$ for various system sizes $M$
  of a box $t$ distribution with $c=0$. Symbols indicate different
  energies ranging from $0.025$ ($\circ$), $0.0225$ ($\square$) to $2
  \times 10^{-5}$ ($\square$). The data right of the broken line have been 
  used in FSS.}
\end{figure}
\begin{figure}[tb]
\vspace{0pt}
\centerline{\psfig{figure=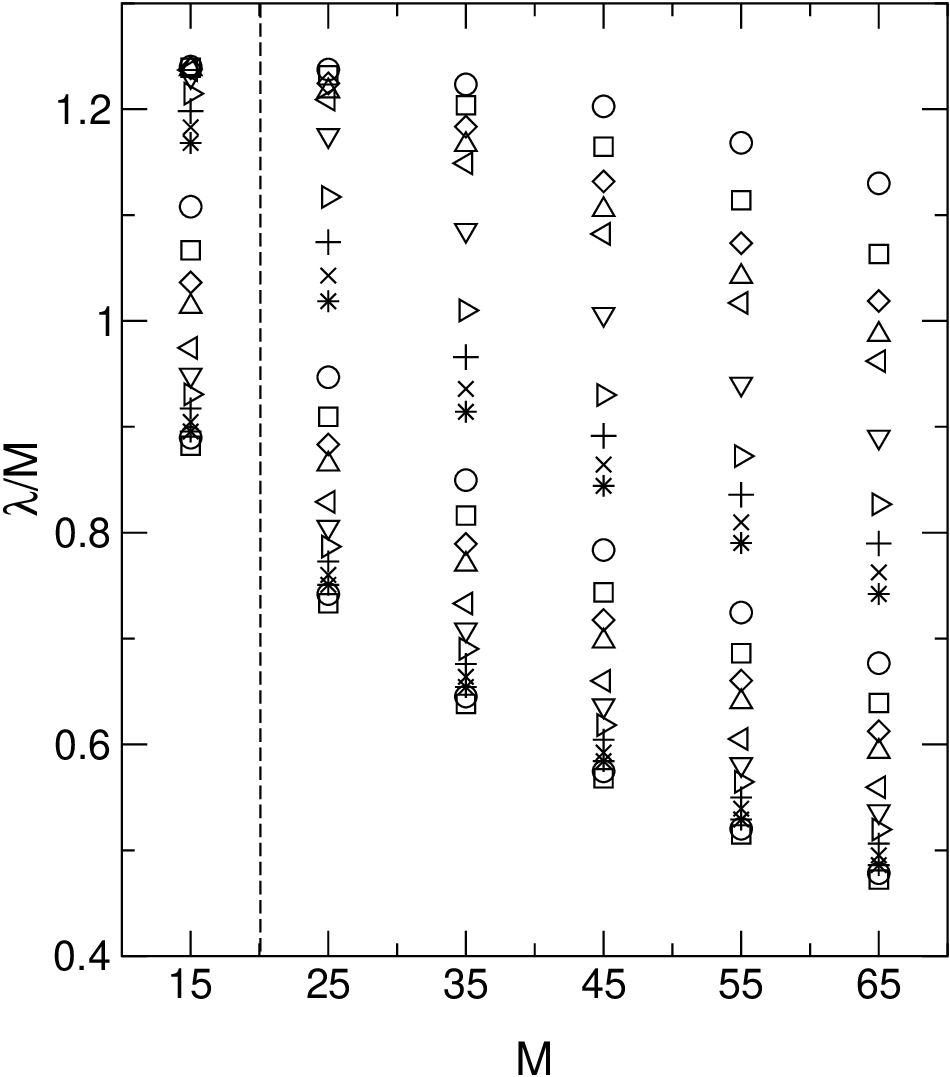,width=0.95\columnwidth}}
\caption{\label{fig-c000Bo}
  Reduced localization length $\lambda/M$ for various odd system sizes
  $M$ of a box $t$ distribution with $c=0$. Symbols indicate different
  energies ranging from $0.025$ ($\circ$), $0.0225$ ($\square$) to $2
  \times 10^{-5}$ ($\square$). The data right of the broken line have
  been used in FSS.}
\end{figure}

We test three different distributions of $t_{ij}$:
\begin{enumerate}
\item a rectangular distribution \cite{EilRS98a}
\begin{displaymath}
P(t_{ij}) = 
\left\{ \begin{array}{ll}
    1/W & \textrm{if $\left|t_{ij}-c\right| \leq w/2$,}\\
    0   & \textrm{otherwise,}
\end{array}\right.
\end{displaymath}
\item a Gaussian distribution
\begin{displaymath}
 P(t_{ij})= \frac{1}{\sqrt{2\pi\sigma^{2}}}
            \exp \left[ -\frac{(t_{ij}-c)^2}{2\sigma
    ^2} \right],    
\end{displaymath}    
\item a rectangular
distribution of the logarithm of $t_{ij}$ \cite{SouWGE82}
\begin{displaymath}
P(\ln t_{ij}/t_0) = \left\{ \begin{array}{ll}
       1/w & \textrm{if $\left| \ln t_{ij}/t_0 \right| \leq w/2$,}\\ 
       0   & \textrm{otherwise.}
        \end{array} \right. 
\end{displaymath} 
\end{enumerate}
The logarithmic distribution appears more suited to model actual
physical systems \cite{SouWGE82}.  We also note that the logarithmic
distribution avoids problems with zero $t$ elements and thus there is
no need to introduce an artificial lower cutoff as for the box and
Gaussian distributions \cite{EilRS98a}.  Furthermore, the box and
Gaussian distributions will usually have negative $t$ values which
correspond to a rather artificial phase shift.

In the case of rectangular and normal distributions we set the width
$w$ and the standard deviation $\sigma$ to $1$ and change the center
$c$ of the distribution. In the case of the logarithmic $t$
distribution $t_0 = 1$ sets the energy scale and we change the
disorder width $w$.  Values of the parameters were $c=0$, $0.25$,
$0.5$, and $1.0$ for the rectangular distribution; $c=0$ and $c=0.25$
for the Gaussian distribution and $w=2$, $6$, and $10$ for the
logarithmic $t$ distribution.

\begin{figure}[t]
\vspace{0pt}
\centerline{\psfig{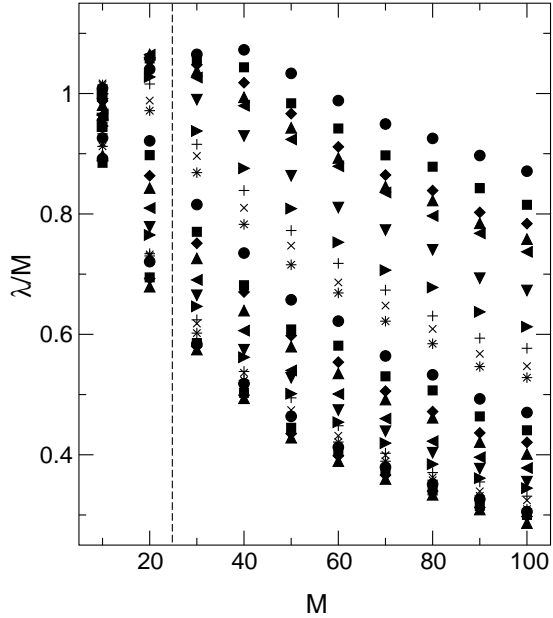}}
\caption{\label{fig-c025G}
  Reduced localization length $\lambda/M$ for various system sizes $M$
  of a Gaussian $t$ distribution with $c=0.25$. Symbols indicate
  different energies ranging from $0.03$ ($\bullet$), $0.0275$
  ($\blacksquare$) to $2 \times 10^{-5}$ ($\blacktriangle$). The data
  right of the broken line have been used in FSS.}
\end{figure}
\begin{figure}[thb]
\vspace{0pt}
\centerline{\psfig{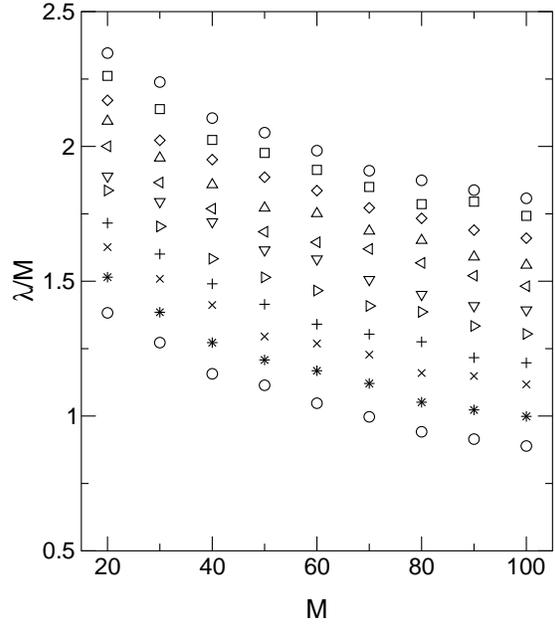}}
\caption{\label{fig-w020L}
  Reduced localization length $\lambda/M$ for various system sizes $M$
  of a logarithmic $t$ distribution with $w=2$. Symbols indicate
  different energies ranging from $0.2048$ ($\circ$), $0.1024$
  ($\square$) to $2 \times 10^{-4}$ ($\circ$).}
\end{figure}

\section{Computation of the localization lengths at $E\neq 0$}
\label{sec-fss}

The transfer-matrix method \cite{MacK81,MacK83} was used to compute
the localization lengths for strips of various widths $M$ up to
$M=100$ in the energy interval $2 \times 10^{-5} \leq E \leq 0.2048$.
In Figs.\ \ref{fig-c000B}, \ref{fig-c000Bo}, \ref{fig-c025G}, 
and \ref{fig-w020L} we show the system size dependence for, e.g., 
special values of $c$ and $w$ and all three disorder distributions.  
The accuracy of our results is $0.1-0.3 \%$ or $1 \%$ depending on 
the disorder distribution and the values of parameters, see Table 
\ref{tab-exp} for actual parameter values. 

\begin{figure}[htb]
\vspace{9pt}
\centerline{\psfig{figure=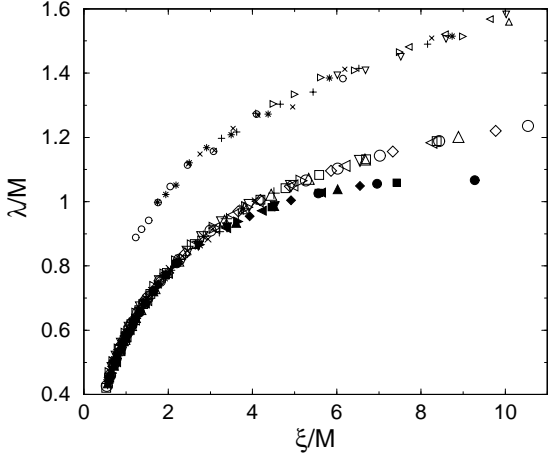,width=0.95\columnwidth}}
\caption{\label{fig-fss}
  Finite-size-scaling plots for box ($c=0$, $M\in [30,80]$, large open
  symbols), Gaussian ($c=0$, $M\in [20,60]$, filled symbols), and
  logarithmic ($w=2$, $M\in[30,100]$, small open symbols)
  $t$ distributions.}
\end{figure}

Next, the finite-size-scaling analysis of Ref.\ \cite{MacK83}
was applied to the data.  The calculated localization lengths usually
increase as the energy approaches $0$. Only, for small even width
values ($10,20$) it decreases significantly close to $E=0$
\cite{BisCRS99} which makes finite-size scaling impossible. Therefore
the smallest system sizes were dropped during the finite-size scaling
procedure. Results for the finite-size scaling curves are shown in
Fig.\ \ref{fig-fss} for the three different distributions.

\section{Critical exponents}
\label{sec-crit-exp}

\begin{figure}[tb]
\vspace{4ex}
\centerline{\psfig{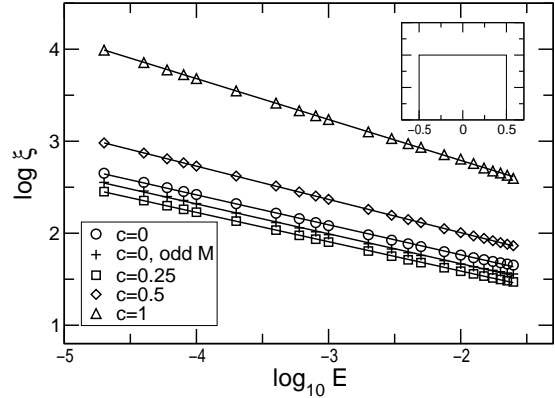}}
\caption{\label{fig-ri}
  Variation of the infinite-size localization length $\xi$ with $E$
  for box distributions. The inset shows the $t$ distribution for
  $c=0$.}
\end{figure}

\begin{figure}[bt]
\vspace{4ex}
\centerline{\psfig{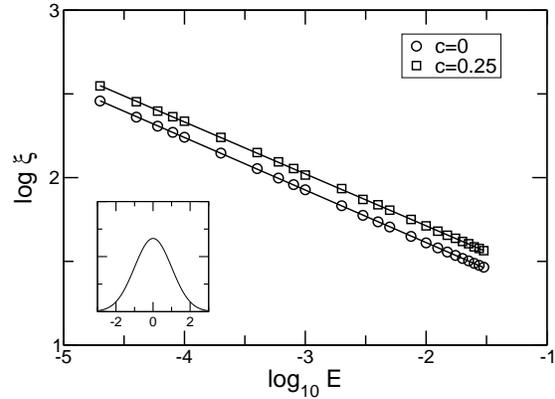}}
\caption{\label{fig-gi}
  Variation of the infinite-size localization length $\xi$ with $E$
  for Gaussian distributions. The inset shows the $t$ distribution for
  $c=0$.}
\end{figure}

\begin{figure}[tb]
\vspace{5ex}
\centerline{\psfig{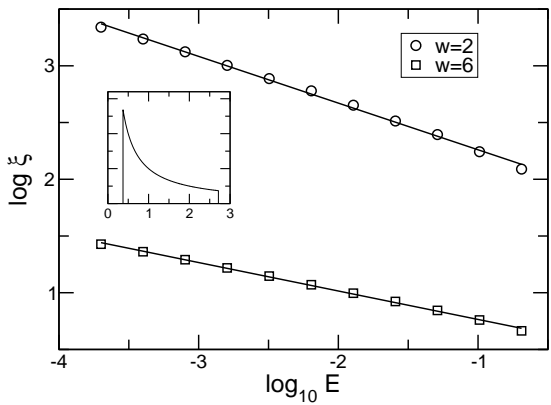}}
\caption{\label{fig-li}
  Variation of the infinite-size localization length $\xi$ with $E$
  for logarithmic distributions. The inset shows the $t$ distribution for
  $w=2$.}
\end{figure}

One expects that the scaling parameters $\xi$ obtained from finite-size scaling
diverge close to $E=0$ \cite{FabC00}. However, the precise functional
form of this divergence is not yet know. In Ref.\ \cite{FabC00} it has
been suggested that for energies $E>E^{*}$ the divergence can be
described by a power law as
\begin{equation}\label{eq-power-law}
   \xi(E) \propto \left| \frac{E_0}{E} \right|^{\nu}
\end{equation}
with the critical exponent $\nu$. For even smaller $|E| \ll E^{*}$,
this behavior should then change to
\begin{equation}\label{eq-exp-law}
  \xi(E) \propto \exp\sqrt{\frac{\ln E_0/E}{A}}
\end{equation}
with constants $E_0$ and $A$ given by the renormalization group flow
\cite{FabC00}.  Double-logarithmic plots of $\xi$ vs.\ $E$ in Figs.\ 
\ref{fig-ri}, \ref{fig-gi} and \ref{fig-li} confirm the power-law
behavior with reasonable accuracy down to $E\approx 10^{-4}$. For
smaller values it has been shown already in Ref.\ \cite{BisCRS99} that
a new behavior is to be expected.

Table \ref{tab-exp} collects the values of the critical exponent
obtained for different disorders. In the case of the logarithmic $t$
distribution and $w=10$ the power-law divergence fails, therefore the
exponent was not calculated.  From Table \ref{tab-exp}, it can be
easily seen that all calculated values are in the range $0.2 \le \nu
\le 0.5$. The exponent is apparently not universal but seems to 
depend on the kind of disorder and the actual value of parameters;
for stronger disorders $\nu$ becomes smaller (for the logarithmic $t$
distribution the disorder strength increases with $w$ \cite{SouWGE82}, for
the rectangular distribution the strongest disorder appears at
$c=0.25$ \cite{EilRS98a}). This non-universality is in agreement with
the results of Ref.\ \cite{FabC00}.
\begin{table*}[htb]
\caption{\label{tab-exp}
  Estimated values of the exponents of the localization lengths for
  various disorder strengths and distributions. The error bars
  represent the standard deviations from the power-law fit and should
  be increased by at least one order of magnitude for a reliable
  representation of the actual errors.}
\newcommand{\m}{\hphantom{$-$}}
\newcommand{\cc}[1]{\multicolumn{1}{c}{#1}}
\renewcommand{\tabcolsep}{2pc} 
\renewcommand{\arraystretch}{1.2} 
\begin{tabular}{@{}lllll}
\hline
disorder     & parameters  & accuracy    & sizes used   & estimated \\
distribution &             & in $\%$     & in finite-size scaling       & exponent \\
\hline
 box         & $c=0$       & $0.1$-$0.2$ &  30-90       & $0.326 \pm 0.002$\\
 box         & $c=0$       & $0.1$-$0.2$ &  25-65       & $0.325 \pm 0.002$\\
 box         & $c=0.25$    & $0.1$-$0.2$ &  30-70       & $0.319 \pm 0.001$\\
 box         & $c=0.5$     & $0.1$-$0.2$ &  30-70       & $0.361 \pm 0.001$\\
 box         & $c=1.0$     & $0.1$-$0.3$ &  30-70       & $0.444 \pm 0.002$\\
 Gaussian    & $c=0$       & $0.1$-$0.2$ &  30-60       & $0.314 \pm 0.001$\\
 Gaussian    & $c=0.25$    & $1$         &  30-100      & $0.310 \pm 0.001$\\
 Gaussian    & $c=0.25$    & $1$         &  35-95       & $0.308 \pm 0.001$\\
 logarithmic & $w=2$       & $1$         &  20-100      & $0.412 \pm 0.007$\\
 logarithmic & $w=6$       & $1$         &  20-100      & $0.251 \pm 0.004$\\
 logarithmic & $w=6$       & $1$         &  25-95       & $0.253 \pm 0.004$\\
 logarithmic & $w=10$      & $1$         &  20-100      & ---            \\
\hline
\end{tabular}
\end{table*}

As the localization lengths calculated for odd and even strip widths
may exhibit different behavior \cite{SouWGE82,BisCRS99} we repeated
the calculations also for odd-width systems for chosen parameter sets 
for all distributions. Fig.\ \ref{fig-c000Bo} shows an example of 
reduced localization lenghts for rectangular distribution. In contrast 
to the even-width systems (Fig.\ \ref{fig-c000B}) the localization 
lengths do not decrease significantly close to the $E=0$.
We have attributed the different behaviors for odd and even system
sizes to different structures in the density of states
\cite{BisCRS99,MilRS99a}.
Nevertheless, the exponent in both cases is within the error bars the
same (cp.\ Table \ref{tab-exp}).  This is also true for Gaussian and
logarithmic disorder distributions (see Table \ref{tab-exp}),
therefore, at least for the investigated disorder strengths, the
difference in exponents of localization lengths is negligible.

\section{Conclusions}
\label{sec-concl}

Our results suggest that the localization-delocalization-localization
present in the off-diagonal Anderson model of localization in 2D can
be described by a set of exponents that model the divergence of the
localization lengths $\xi$ at $E=0$. Note that these exponents are in
reasonable agreement with the exponent $0.5$ first estimated for the
scaling of the participation numbers in Ref.\ \cite{EilRS98a}.
Down to $E\approx 2 \times 10^{-5}$ in Figs.\ \ref{fig-ri}, \ref{fig-gi}
the power-law behavior can model the data reasonably well. Thus we 
expect the crossover predicted in Ref.\ \cite{FabC00} to appear at 
smaller energies. We find that the exponents depend on the strength and 
distribution of the off-diagonal disorder also in agreement with 
Ref.\ \cite{FabC00}. Currently, we are extending these calculations 
to smaller energies.

We note that it might be interesting to also investigate the situation
in honeycomb lattices \cite{SchO91}, where the van Hove singularity of
the square lattice at $E=0$ does not interfere with the divergence due
to the bipartiteness which is of interest here.

\section{Acknowledgments}

We thank M.\ Fabrizio for stimulating discussions.  We gratefully
acknowledge support by the Deutsche Forschungsgemeinschaft (SFB393)
and the SMWK.

%
%


\end{document}